\documentclass[11pt]{article}
\usepackage{graphicx,verbatim,array,multicol,palatino,amssymb,amsfonts, amsmath, subfigure, epsfig}
\usepackage{color}
\usepackage{psfrag}
\usepackage{chicago}

\def\bx{{\bf x}}

\def\lboxit#1{\vbox{\hrule\hbox{\vrule\kern6pt
      \vbox{\kern6pt#1\kern6pt}\kern6pt\vrule}\hrule}}

\def\thick#1{\hbox{\rlap{$#1$}\kern0.25pt\rlap{$#1$}\kern0.25pt$#1$}}

\def\bbeta{{\thick\beta}}
\def\btheta{{\thick\theta}}
\def\bnu{{\thick\nu}}
\def\bxi{{\thick\xi}}

\def\bgamma{{\thick\gamma}}

\def\bSig{{\bf \Sigma}}

\def\bx{{\bf x}}

\def\bS{{\bf S}}
\def\bU{{\bf U}}
\def\bxi{{\bf \xi}}

\def\jump{\vskip3mm\noindent}

\begin{document}
\null\vskip2cm
\begin{center}

{\LARGE
\bf Approximate Bayesian Computation via Regression Density Estimation}
\vskip5mm
\begin{large}

{\sc Y. Fan},\hspace{-1mm}
\footnote{School of Mathematics and Statistics, University of New South Wales, Sydney, 2052, AUSTRALIA}
\quad
{\sc D. J. Nott}\hspace{-1mm}
\footnote{Dept. of  Statistics and Applied Probability, National University of Singapore, SINGAPORE, 117546}
\quad and \quad
{\sc S. A. Sisson}$^1$

\end{large}
\vskip5mm

{\sc Abstract}\vskip2mm
\end{center}
\jump
Approximate Bayesian computation (ABC) methods, which are applicable when the likelihood
is difficult or impossible to calculate, are an active topic of current research.  Most current
ABC algorithms directly approximate the posterior distribution, but an alternative, less common strategy 
is to approximate the likelihood function.  This has several advantages.
First, in some problems, it is easier to approximate the likelihood than to approximate the
posterior.  Second, an approximation to the likelihood allows reference analyses to be constructed
based solely on the likelihood.   Third, it is straightforward to perform sensitivity analyses for 
several different choices of prior once an approximation to the likelihood is constructed, which needs
to be done only once.  The contribution of the present paper is to consider regression density estimation techniques
to approximate the likelihood in the ABC setting.  Our likelihood approximations 
build on recently developed marginal adaptation density estimators by extending them for 
conditional density estimation.  Our approach facilitates reference Bayesian inference, as well as 
frequentist inference. The method is demonstrated via a challenging problem of inference 
for stereological extremes, where we perform both frequentist and Bayesian inference.  \\

\noindent
{\em Keywords}: Approximate Bayesian computation; Copulas; Likelihood-free inference; Mutivariate density estimation; Regression density estimation.
\vskip4mm

\section{Introduction}

Approximate Bayesian computation (ABC) methods (commonly described as ``likelihood-free'' methods) 
have been attracting increasing research interest as a viable procedure for performing 
Bayesian inference in the presence of computationally intractable likelihood functions. Initially popular 
in the biological sciences \shortcite{beaumont+zb02,luciani+sjft09,ratmann+jhsrw07,ratmann+ahwr09}, 
ABC has now found application in a wide range 
of areas -- see the reviews by \shortciteN{beaumont10}, \shortciteN{csillery+bgf10} and \shortciteN{sisson+f11}
\\

ABC methods require the ability to simulate data from the intractable likelihood, 
$L(\cdot | \btheta)$, for a given parameter vector $\btheta$. An integral part of all 
existing algorithms involves 
sampling $\btheta$ from some distribution (typically the prior, $\pi(\btheta)$) and comparing 
simulated data $\bx\sim L(\cdot|\btheta)$ with the observed data $\bx_0$.  
Posterior samples approximately from $\pi(\btheta | \bx_0)$ are then obtained by  weighting $\btheta$
according to some weighting function or kernel $K_{\delta}(\|\bx-\bx_0\|)$ with scale parameter $\delta\geq 0$. Accordingly, samples $\btheta$ for which $\bx\approx\bx_0$ receive larger weights than those where $\bx$ is very different to $\bx_0$.
\\


More formally,  ABC algorithms produce samples from an augmented posterior distribution 
\begin{equation}\label{eqnABC}
\pi_{ABC} (\btheta, \bx | \bx_0)\propto K_{\delta}(\|\bx-\bx_0\|) L(\bx|\btheta)\pi(\btheta).
\end{equation}
It is easy to see that as $\delta \rightarrow 0$, only $\bx \approx \bx_0$ is given non-trivial weight by $K_\delta$, and the ABC posterior converges 
to the true posterior distribution, $\pi(\btheta|\bx_0)$. However,  $\delta = 0$ practically
equates to repeated simulation until $\bx = \bx_0$ is matched exactly for each $\btheta$, 
which is computationally impractical in general.
As such, for some $\delta>0$, a necessarily approximate posterior is obtained.
Regression-adjustment strategies have been proposed to post-process the approximate posterior samples to 
correct the error in the ABC posterior (e.g. \shortciteNP{beaumont+zb02,blum+f09,nott+fms11}).
Dimension reduction techniques are also typically used to reduce the overall computational overhead. Here, the vectors, $\bx$, are replaced by lower-dimensional sufficient or summary statistics, ${\bf S} \equiv \bS(\bx)$.
Readers should refer to \shortciteN{blum12}, \shortciteN{prangle12} and
\shortciteN{robert11} for further discussion of the use and elicitation of summary statistics in ABC.\\

In this paper, we present an alternative ABC approach based on constructing a direct approximation to the likelihood. 
The approximation is based on samples $\btheta^{(i)}$ (not necessarily from the prior), and the corresponding 
summaries $\bS^{(i)}$, simulated from the intractable likelihood.
Such an approach is attractive in several ways.  Firstly, it is sometimes easier to approximate the likelihood
than to approximate the posterior directly.  Secondly, separate estimation of the likelihood is
useful for performing purely likelihood based analyses, such as maximum likelihood estimation. This may be more attractive to researchers more familiar with frequentist inference.
Thirdly, direct approximation of the likelihood is useful for diagnosing prior-likelihood conflict, 
%
%
or for running many analyses with different priors in a sensitivity
analysis, as the likelihood approximation only needs to be constructed once.
A further example of a Bayesian use for a separate estimate of the likelihood is the construction
of credible regions based on contours of the likelihood -- these have a robust Bayes interpretation in terms of 
posterior probability content being minimally sensitive to perturbations of the
prior \shortcite{wasserman89}. The method we propose here provides an 
explicit analytical expression for the likelihood function. It also does not require simulation from the prior, which is typically required by many ABC algorithms (see e.g.  \shortciteNP{pritchard+spf99,tavare+bgd97,sisson+ft07}). This is useful where there is interest in using reference priors. \\

An early discussion of the concept of directly approximating an intractable 
likelihood function using simulated data sets was given by \shortciteN{diggle84}, predating
the current wave of activity in ABC research.  More recently, several researchers have considered direct likelihood approximations within ABC.
\shortciteN{leuenberger+we10} 
developed an approach that involves embedding the regression adjustment
of \shortciteN{beaumont+zb02} into a likelihood approximation based on a generalised linear model.  
\shortciteN{wood10} considers a ``synthetic likelihood'' which involves estimating a normal
distribution for the summary statistics, ${\bf S}$, with mean and covariance depending on $\btheta$.  The approximation of the mean and covariance is performed within
a Markov chain Monte Carlo (MCMC) scheme, and through clever choice of summary statistics and
quantile transformations it may be possible to improve the approximation to normality.  
The approach we propose is also related to that introduced by
\shortciteN{bonassi11}, although this method does not directly approximate the likelihood.  
They propose to use mixtures of multivariate normals to estimate a joint distribution for $(\btheta,\bS)$ where the summary statistics are simulated from an informative prior, and then condition on the observed summary statistic
in the estimated mixture model.  Appropriate localization and choice of summary statistics can
help to improve the efficiency of the approach.
\\

We note that a direct approximation of the likelihood function is available by Monte Carlo integration (e.g. \shortciteNP{sisson+pfb08,sisson+f11}).
If we consider marginalising the distribution of the augmented posterior (\ref{eqnABC}) with respect to the 
auxiliary data $\bx$,  we then have the marginal posterior
\begin{equation}\label{mcL}
\pi_{ABC} (\btheta | \bx_0)  \propto \pi(\btheta)\int K_{\delta}(\|\bx-\bx_0\|) L(\bx|\btheta)d\bx 
\approx \frac{\pi(\btheta)}{n}\sum_{i=1}^n K_{\delta}(\|\bx^{(i)}-\bx_0\|)
\end{equation}
where $\bx^{(1)},\ldots,\bx^{(n)}$, are samples from $L(\bx|\btheta)$.
The summation term is a simple Monte Carlo estimate of the likelihood of $\bx_0$ at the point $\btheta$. 
Clearly, this estimate is a function of $\delta$, and so posterior inferences will still require regression-adjustment post-processing. It is also a pointwise estimate, and so it must be effectively re-estimated for each likelihood evaluation in an analysis.
Our approach results in a stand-alone functional expression for the likelihood function at $\delta=0$.
However, in the context of this paper, the estimator  (\ref{mcL}) could usefully serve as a 
goodness-of-fit diagnostic.
\\

Our proposed approach to conditional density estimation for ABC (that is, estimating the distribution of $\btheta|\bS$ given the joint samples $(\btheta,\bS)$) 
builds on recent work on flexible multivariate
density estimation -- in particular the marginal adaptation method of \shortciteN{giordani+mk09}.  
The basis of their approach argues that estimation of univariate marginal distributions is easier
than estimating a full multivariate density estimation directly. Hence, a fruitful strategy is to adapt a multivariate
density estimate to have given marginals, which are more precisely estimated individually.  In this paper, we develop a strategy for implementing this
approach for the related problem of conditional density estimation, in the context of  likelihood estimation in ABC. 
\\

In Section \ref{secRDE}, we develop the estimator for the intractable likelihood function, and Section \ref{secConnect} outlines
some connections with existing ABC approaches. Section \ref{secExtremes} discusses a challenging example of inference
for stereological extremes and Section \ref{secDiscussion} concludes.

\section{ABC via multivariate regression density estimation}
\label{secRDE}

Our goal is to obtain an estimate
 of  the likelihood $L(\bS| \btheta)$, denoted by $\hat{L}(\bS| \btheta)$, and then approximate the intractable posterior distribution by
$$
\hat{\pi}(\btheta | \bS_0) \propto \hat{L}(\bS_0| \btheta)\pi(\btheta),
$$%
where $\bS_0=\bS(\bx_0)$ denotes the summary statistics from the observed data. Once $\hat{L}(\bS_0|\btheta)$ is obtained, drawing samples from $\hat{\pi}(\btheta|\bS_0)$  can be achieved by any standard posterior simulation algorithm e.g. Markov chain Monte Carlo. 
We assume that the vector of summary statistics, $\bS=(S^1,\dots,S^k)^\top$, are given, with $\dim(\bS)=k$ and that $\dim(\btheta)=d$.  
\\

The likelihood estimate is constructed from samples $(\bS^{(i)},\btheta^{(i)})$ for $i=1,\ldots,N$, where $\btheta^{(i)}\sim h(\btheta)$ and $\bS^{(i)}=\bS(\bx^{(i)})$ with $\bx^{(i)}\sim L(\cdot|\btheta^{(i)})$. The distribution function $h(\btheta)$ determines the region of parameter space in which we require $\hat{L}(\bS_0|\btheta)$ to be a good approximation of $L(\bS_0|\btheta)$. 
This must necessarily include the region of high posterior density.
One option, where this is convenient, is to perform an initial pilot ABC analysis with a large value of $\delta$ to broadly identify the high posterior density region, and then define $h(\btheta)$ to be proportional to the prior $\pi(\btheta)$ in this region \cite{prangle12}.
\\

We first consider flexible estimation of the
individual marginal distributions $S^j|\btheta$ for each $j=1,\ldots, k$. Estimation of marginals is 
typically easier than estimation of the joint distribution due to the lower dimensionality, and there are many different methods available for 
this purpose. Here we adopt a mixture of experts approach \shortcite{jacobs+jnh91,jordan+ja94}.  
Mixtures of experts models are mixtures of regression models where both parameters in the component
response distributions and mixing weights are allowed to vary with covariates.  
In our case, for each $S^j$, we express the marginal distribution of $S^j|\btheta$ as a mixture of $J$ normal distributions
\begin{equation}\label{eqnS}
f_j(S^j |\btheta)  = \sum_{\ell=1}^J w_{j\ell} (\btheta) N\left(\mu_{j\ell}(\btheta), \sigma^2_{j\ell}(\btheta)\right)
\end{equation}
where
$$ w_{j\ell}(\btheta) = \frac{\exp\left(\xi_0^{j\ell}+ {\bxi^{j\ell}}^\top\btheta\right)}
{\sum_{k=1}^J \exp\left(\xi_0^{jk}+ {\bxi^{jk}}^\top\btheta\right)}.
$$
Here $\xi_0^{j\ell}\in \mathbb{R}$ and $\bxi^{j\ell}\in \mathbb{R}^d$, for $j=1,\dots k$, $\ell=1,\dots J$ with
$\xi_0^{j1}=0, \bxi^{j1}={\bf 0}$ for identifiability, and
$$
\mu_{j\ell} (\btheta) = \beta_0^{j\ell}+ {\bbeta^{j\ell}}^\top\btheta, \quad \log\left(\sigma^2_{j\ell} (\btheta)\right) = \gamma_0^{j\ell}+ {\bgamma^{j\ell}}^\top\btheta ,
$$
with $\beta_0^{j\ell}, \gamma_0^{j\ell} \in R$ and $\bbeta^{j\ell}, \bgamma^{j\ell} \in R^d$.
For a sufficiently large sample size ($N$) and number of components ($J$),  the model (\ref{eqnS}) can approximate the 
marginal distribution of $S^j| \btheta$ arbitrarily well. 
Methods for fast and parsimonious fitting of  (\ref{eqnS}) can be found in 
\shortciteN{notttvk11} and \shortciteN{trannk12}.\\

For the estimation of the joint dependencies, we propose a mixture of Gaussian copulas approach
similar to that proposed by \shortciteN{giordani+mk09} for density estimation.  More
precisely, for joint samples of $(\bS^{(i)},\btheta^{(i)})$ 
we firstly transform the fitted margins of $S^j$ to be standard normal, so that
\begin{equation}\label{Uj}
U^j= \Phi^{-1}(F_j(S^j|\btheta)),
\end{equation}
where $\bU=(U^1,\ldots,U^k)$, $F_j$ denotes the cumulative distribution function for the fitted density $f_j(\cdot)$ in (\ref{eqnS})
and $\Phi$ denotes the standard normal cumulative distribution function.
We then fit a mixture of multivariate normals to the joint samples 
$(\bU, \btheta)$, and obtain the conditional distribution of $\bU | \btheta$
in this mixture (see e.g. \shortciteNP{noretsp11}). In more detail, the joint density of $(\bU,\btheta)$ is modelled via a mixture of multivariate normal distributions
\begin{equation}\label{eqnJoint2}
g(\bU, \btheta) = \sum_{\ell=1}^L \omega_\ell N_{k+d}(\bnu_\ell,\bSig_\ell),
\end{equation}
where 
$\sum_{\ell=1}^L \omega_\ell =1$ and
$\bnu_\ell, \bSig_\ell$ denote the mean and covariance matrices of the normal mixture components. 
The conditional density
$g(\bU| \btheta)$ is then given by
\begin{equation}\label{eqnCond}
g(\bU|\btheta) =  \sum_{l=1}^L \omega^c_\ell N_{k}(\bnu^c_\ell,\bSig^c_\ell),
\end{equation}
where $\bnu^c_\ell$ and $\bSig^c_\ell$ are the mean and covariance matrix of the conditional distribution 
given $\btheta$ in the the $\ell$-th mixture component, $N_{k+d}(\bnu_\ell,\bSig_\ell)$.  
The mixing weights for the conditional distribution, $\omega^c_\ell$, are 
$$
\omega^c_\ell = \frac{\omega_\ell \phi\left(\btheta, \bnu_\ell(\btheta), \bSig_\ell(\btheta)\right)}{\sum_{i=1}^{L} \omega_i\phi\left(\btheta, \bnu_i(\btheta), \bSig_i(\btheta)\right)}
$$
where $\phi\left(\btheta, \bnu_\ell(\btheta), \bSig_\ell(\btheta)\right)$ denotes the marginal density for $\btheta$ 
in the normal component $N_{k+d}(\bnu_\ell,\bSig_\ell)$.  \\

Our density estimator for the likelihood is then given as
\begin{equation}\label{eqn:likest}
\hat{L}(\bS| \btheta) = g(\bU|\btheta)\prod_{j=1}^k \frac{f_j(S^j|\btheta)}{\phi(U^j, 0, 1)},
\end{equation}
where $\phi(U^j,0,1)$ denotes a standard normal density evaluated at $U^j$.  
Note that $\hat{L}(\bS| \btheta)$ does not have exactly $f_j(S^j|\btheta)$ as it's marginal distribution, since the marginal distribution for $U^j$ under $g(\bU|\theta)$ is not exactly
standard normal, unless $L=1$.  
For this reason, \shortciteN{giordani+mk09} suggest recomputing the distribution of $U^j$ using another normal
mixture, which would lead to replacing $\phi(U^j,0,1)$ with this distribution in (\ref{eqn:likest}). This will improve the quality of the likelihood approximation, albeit at additional computational cost. However in our experience we have found that an acceptable approximation is obtained even without enforcing the marginals $f_j(S^j|\btheta)$ exactly.
\\

A key component of the above approach is the transformation $(\bS,\btheta)\rightarrow(\bU,\btheta)$. If the summary statistics are too highly correlated, or if the dependence between the statistics and the parameters is too complex, it will be difficult to capture the joint distribution of $(\bS,\btheta)$ well in a simple mixture model. The primary role of the flexible mixture of experts approach (\ref{eqnS}) is to construct the transformation $(\bS,\btheta)\rightarrow(\bU,\btheta)$ to  simplify this complexity. However, such modelling can be greatly aided by careful initial selection of summary statistics to achieve near orthogonality and low dependence (e.g. \shortciteNP{blum12}, \shortciteNP{prangle12}). 
\\

A step by step description of how to of construct $\hat{L}(\bS|\btheta)$ is:

\begin{description}

\item[Step 1]  Sample $(\bS^{(i)},\btheta^{(i)})\sim L(\bS|\btheta)h(\btheta)$ for $i=1,\ldots,N$.
Define $\bS^{(i)}=\left(S^{(i)1},\ldots,S^{(i)k}\right)$.

\item[Step 2] For each $S^j$, fit the mixture model (\ref{eqnS}) to the marginal samples $(S^{(i)j},\btheta^{(i)})$.

\item[Step 3] Compute the transformations $U^{(i)j}=\Phi^{-1}\left(F_j(S^{(i)j}|\btheta)\right)$, for $j=1,\ldots,k$, 
$i=1,\ldots,N$.   Define $\bU^{(i)}=\left(U^{(i)1},\ldots,U^{(i)k}\right)$.

\item[Step 4] Fit the mixture model (\ref{eqnJoint2}) to the samples $(\bU^{(1)},\btheta^{(1)}),\ldots, (\bU^{(N)}, \btheta^{(N)})$ 
and derive the model $g(\bU| \btheta) $ from (\ref{eqnCond}).

\item[Step 5] Combine the results in Steps 2, 3 and 4 to obtain the final estimate for the likelihood
$\hat{L}(\bS|\btheta)$ via (\ref{eqn:likest}).


\end{description}

The estimate, $\hat{L}(\bS|\btheta)$, can then be used as part of any standard Bayesian or frequentist analysis.


\section{Connections with other methods}
\label{secConnect}

\shortciteN{bonassi11} provide a method to directly estimate the posterior density. A mixture of multivariate normals is used to approximate the joint distribution of $(\bS,\btheta)$ based on samples $(\bS^{(i)},\btheta^{(i)})\sim L(\bS|\btheta)
\pi(\btheta)$, where the prior must be informative. Conditioning this mixture on $\bS=\bS_0$ then provides an estimate of the posterior. To the best of our knowledge, this is the only approach that directly uses non-parametric conditional density estimation techniques in ABC. Note that alternatively conditioning the mixture approximation of $(\bS,\btheta)$ on $\btheta$ would result in a mixture of normals conditional density estimate for the likelihood. Similarly, we may derive a direct estimate of the posterior similar to (\ref{eqn:likest}) by conditioning on $\bS$ rather than $\btheta$, although the sampling distribution $h(\btheta)$ must then equal the prior $\pi(\btheta)$. Arguments for estimating likelihoods, rather than posterior distributions are provided in the Introduction.
\\

\shortciteN{giordani+mk09} discuss the performance of the mixture of multivariate
normals density estimator, and observe that it can be difficult to estimate the 
marginals well using this approach.
Our conditional estimator (\ref{eqn:likest}) uses a mixture of normals to estimate 
the joint distribution of transformed data.  The benefit of the transformation $(\bS,\btheta)\rightarrow(\bU,\btheta)$ is that it is simpler to estimate the joint distribution of $(\bU,\btheta)$ than that of $(\bS,\btheta)$. The mixture of normals model also provides greater flexibility over Gaussian copula models which would simply fit a normal distribution to the transformed data.  
\\

There are clear benefits in using more flexible density estimation methods. For example, the flexibility in the mixture of experts model (\ref{eqnS}) to describe the relationship between $S^j$ and $\btheta$, means that good transformations $(\bS,\btheta)\rightarrow(\bU,\btheta)$ can be constructed for a wide range of $\btheta$ vectors drawn from $h(\btheta)$. As such, we can obtain good approximations to $L(\bS|\btheta)$ for a wide range of $\btheta$ as long as the regression models in (\ref{eqnS}) are adequate and enough samples $(\bS^{(i)},\btheta^{(i)})$ are available. Clearly, well-chosen summary statistics can also make the modelling easier, especially when they behave nearly linearly with the parameters. See  \shortciteN{blum12} for a review of dimension reduction methods for ABC.
\\

Finally, we note that if the initial sample $(\bS^{(i)},\btheta^{(i)})$ is obtained via a pilot ABC analysis with a moderately large kernel scale parameter $\delta$, then our approach can be viewed as another form of ABC posterior adjustment method. This is in the sense that the initial samples are used to estimate the likelihood at the point $\delta=0$, from which (new) samples are subsequently obtained. See e.g.  \shortciteN{leuenberger+we10} and other regression adjustment postprocessing methods \shortcite{beaumont+zb02,blum+f09,nott+fms11} for further information.

\section{Example: Inference for stereological extremes}%
\label{secExtremes}

We now present a reanalysis of a stereological problem studied by \shortciteN{bortot+cs07} focusing on the production of clean steels. Here inference is required on the size and intensity of the largest microscopic particle inclusions in a 3-dimensional block of steel, based on inclusion cross sections observed in a 2-dimensional planar slice.
The observed data consist of 112 inclusions intersecting the planar slice, with cross sectional diameters above a measurement threshold of $\nu_0=5\mu m$.
We follow \shortciteN{bortot+cs07} and consider two models. The first assumes a Poisson process with rate $\lambda$ for inclusion centres, and a size distribution for the inclusions, which are assumed to be spherical, based on univariate extreme value theory. Here, conditional on exceeding the threshold, $\nu_0$, inclusion sizes are assumed to follow a generalised Pareto distribution with scale and shape parameters $\sigma>0$ and $\xi$. In this setting it is possible to perform standard Bayesian inference, and so the correct posterior distribution is available. The second model is the same as the first, 
but assumes that the inclusions are ellipsoidal. For each model we have $\btheta=(\lambda,\sigma,\xi)$.
\\

The analysis of \shortciteN{bortot+cs07} took the number of inclusions, and 112 equally spaced quantiles, $q_{(1)},\ldots,q_{(112)}$, as summary statistics. Here, as the quantiles will be highly correlated, we instead use $S^j=\log(q_{(j+1)}-q_{(j)})$ for $j=1,\ldots, 111$, where $S^{112}$ is the log number of inclusions.
In order to determine the distribution $h(\btheta)$ in regions of non-negligible posterior density, we
found it convenient to first draw samples from $I(\|\bS-\bS_0\|\leq\delta)L(\bS|\btheta)U(\btheta)$, where $\|\cdot\|$ denotes squared Euclidean distance, for a large value of $\delta=20$, where $U(\btheta)$ is uniform over a moderately large region of parameter space expected to contain the posterior.
We then defined 
$$h(\btheta)= N(\hat{\mu},\hat{\Sigma})I\left((\btheta-\hat{\mu})^\top\hat{\Sigma}^{-1}(\btheta-\hat{\mu})\leq3d\right)$$
 as a truncated normal distribution, where $\hat{\mu}$ and $\hat{\Sigma}$ denote the mean and covariance of the $\btheta$-component of these samples. In this manner, we obtained a final sample $(\bS^{(i)},\btheta^{(i)})\sim L(\bS|\btheta)h(\btheta)$ of size 5,000 for each model.
\\

In the following, we demonstrate the performance of the regression density estimation methodology using the 
above summary statistics (corresponding to conditional density estimation in 115 dimensions), as well as the ``semi-automatic'' summary statistics of \shortciteN{prangle12}, which provide one summary statistic per parameter (i.e. 6-dimensional density estimation). In addition to standard ABC posterior inference we also perform a prior sensitivity analysis and a maximum-likelihood based inference.

\subsection{Results}%

\begin{figure}
\begin{center}
\includegraphics[width=15cm]{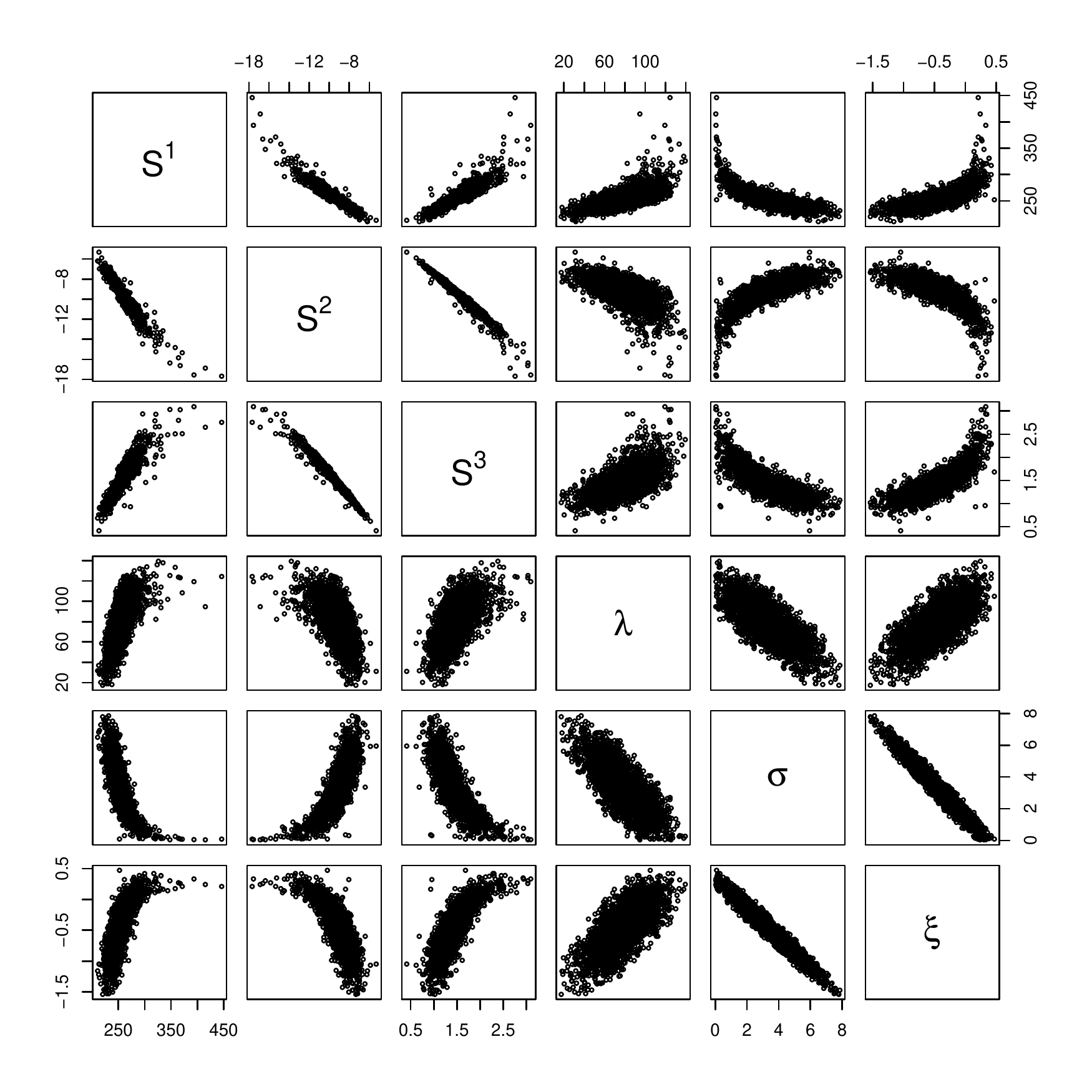}
\end{center}
\caption{  \label{pairsStheta}
Pairwise scatterplots between the three semi-automatic summary statistics $S^1, S^2, S^3$ (Fearnhead and Prangle, 2012) and the three parameters $\lambda, \sigma, \xi$ for the ellipsoidal inclusions model. }
\end{figure}

\begin{figure}
\begin{center}
\includegraphics[width=15cm]{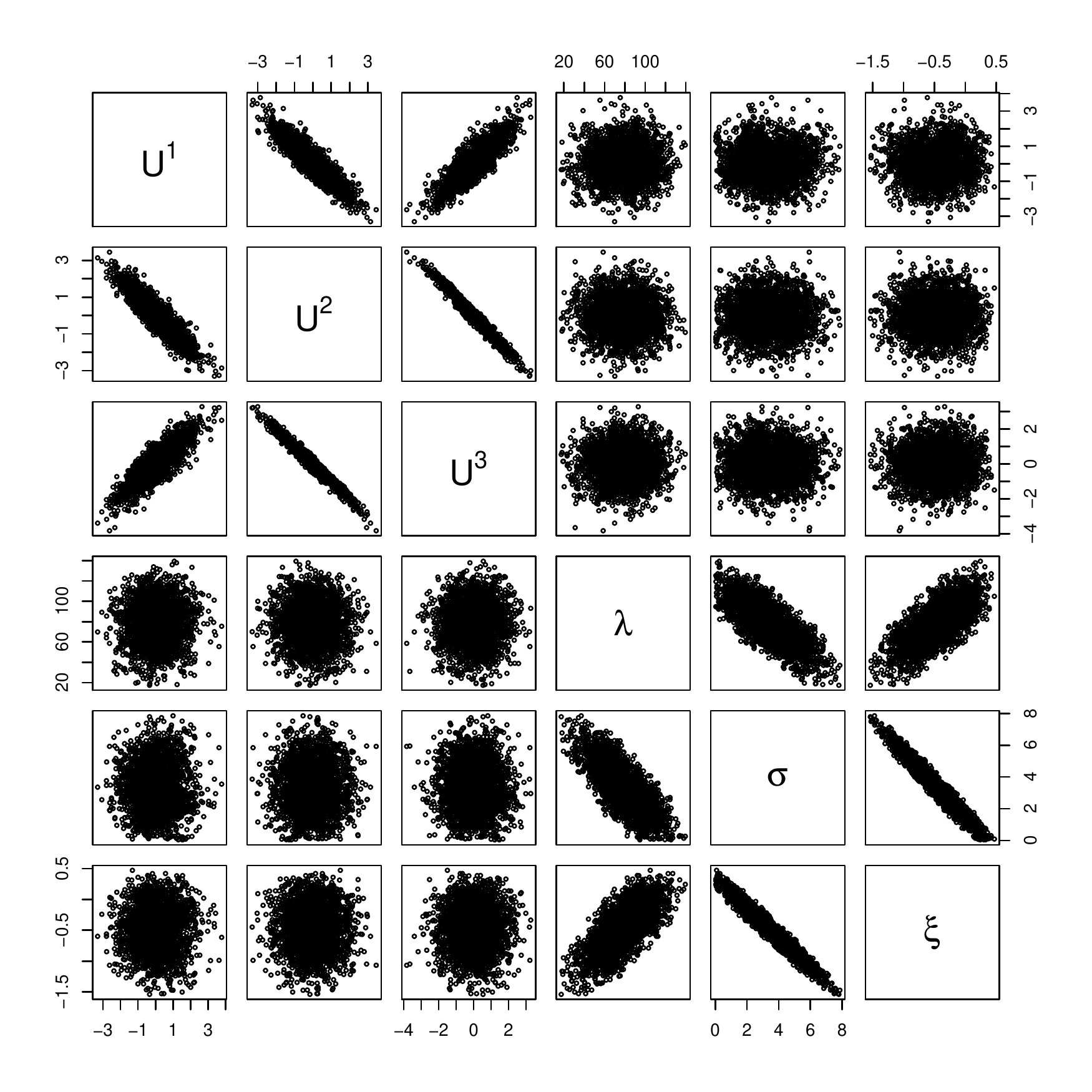}
\end{center}
\caption{  \label{pairsUtheta}
Pairwise scatterplots between the mixture of experts transformed statistics, $U^1, U^2, U^3$ and the three parameters $\lambda,\sigma,\xi$ for the ellipsoidal inclusions model.}
\end{figure}

To fit the mixture of experts marginal distributions  $f_j$ (\ref{eqnS}), we adopted the variational Bayes 
approach of  \shortciteN{notttvk11} and used the variational lower bound to select the number of mixture components.  
For both spherical and ellipsoidal models we found $J=3$ mixture components worked well for each of the 112 summaries, although $J=6$ components were required for the 3 semi-automatic statistics.   
Note that the fitting of the marginal models, $f_j$, is highly parallelizable, and is accordingly suitable for applications in high dimensions.
 Figure \ref{pairsStheta} 
shows the relationship between the parameters $(\lambda,\sigma,\xi)$ and the three semi-automatic summary statistics, as observed in the sample $(\bS^{(i)},\btheta^{(i)})\sim L(\bS|\btheta)h(\btheta)$ for the ellipsoidal model.
There are visibly strong relationships both within the summary statistic and parameter vectors, and between them. Figure \ref{pairsUtheta} displays the same information following the transformation $(\bS,\btheta)\rightarrow(\bU,\btheta)$. Clearly, the transformation has greatly simplified the underlying dependence structure. As a result, while the joint distribution of $(\bS,\btheta)$ could be reasonably well modelled by a mixture of multivariate normal distributions (e.g. \shortciteNP{bonassi11}), a better fit is likely to be obtained by alternatively fitting to $(\bU,\btheta)$.
\\

\begin{figure}
\begin{center}
\includegraphics[width=15cm]{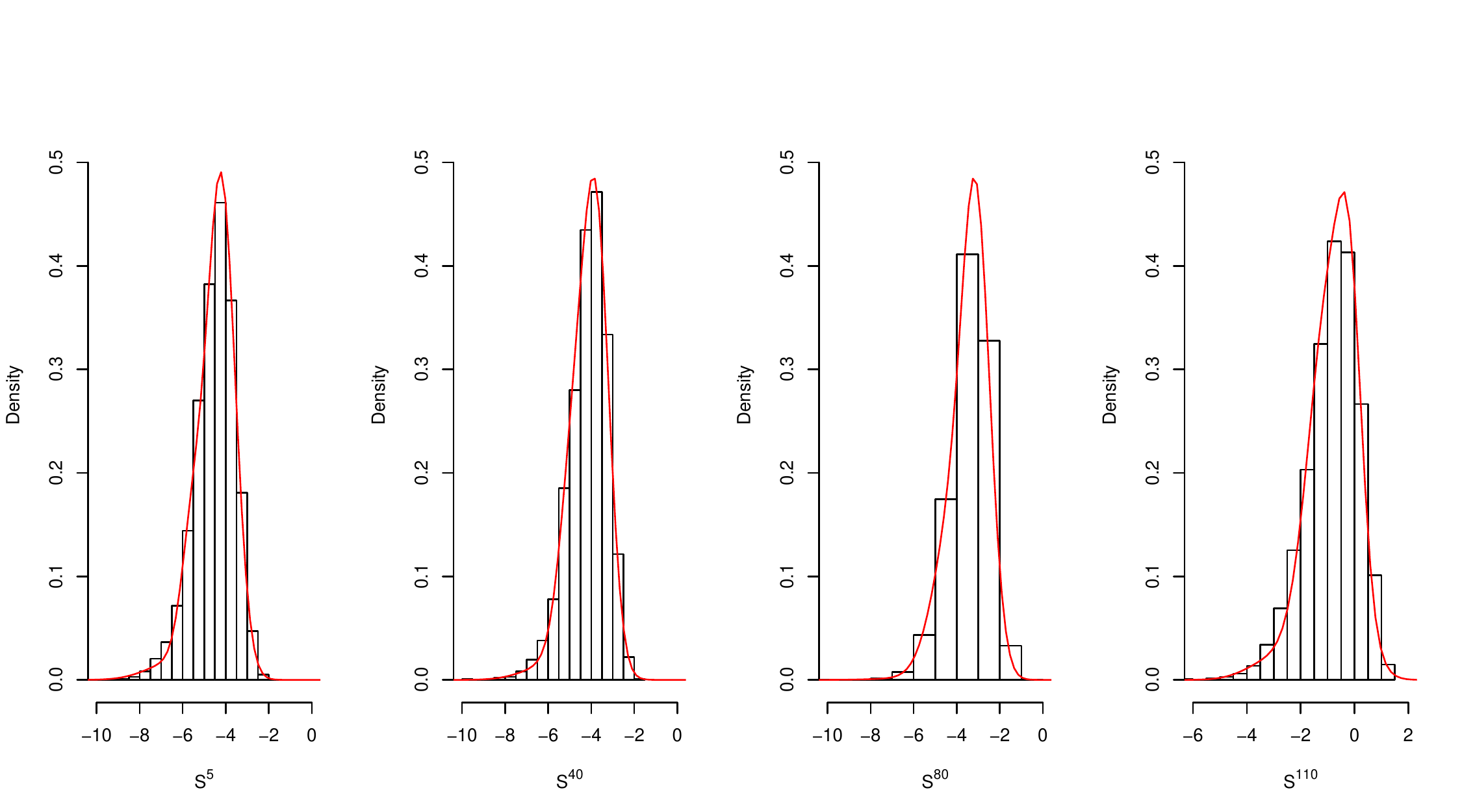}
\end{center}
\caption{  \label{ME-sphere}
Mixture of experts model fits of $f_j(S^j|\btheta)$ for $j=5, 40, 80$ and $110$. Solid line indicates fitted density at the point $\btheta_0=(\lambda_0,\sigma_0,\xi_0)=(30,1.5,-0.05)$. Histogram denotes the true density obtained by drawing samples from $L(S^j|\btheta_0)$ at the point $\btheta_0$. 
}
\end{figure}

Figure \ref{ME-sphere} shows the fitted marginal density estimate, $f_j(S^j|\btheta)$, for $S^j, j=5, 40, 80, 110$ under the spherical model, evaluated at $\btheta_0=(\lambda_0,\sigma_0,\xi_0)=(30,1.5,-0.05)$. The histogram corresponds to the true density, obtained from data generated under the model $L(\bS|\btheta_0)$ at the point $\btheta_0$, and the solid line denotes the fitted density.
The plots indicate a very good fit, and this was typical of all other summary statistics, $S^j$, and conditioned parameters, $\btheta_0$, examined.
These results indicate that marginal densities can be very well estimated. Indeed, marginal densities are much easier to estimate in general than joint densities. \\

We used the {\em R} package {\tt Mclust} to fit the joint distribution $g(\bU,\btheta)$ (\ref{eqnJoint2}) to the transformed samples, using $L=5$ mixture components for both spherical and ellipsoidal models.
Model selection for the number of mixture components is typically performed using the BIC. However care should be taken in high dimensions, 
as the dimension of $(\bU,\btheta)$ can grow quickly, and the BIC tends to overpenalise the number of parameters. 
\shortciteN{giordani+mk09} use the variational Bayes approximation method to fit the mixture model, 
and advocate the use of the variational lower bound for model selection.  
\\

\begin{figure}
\begin{center}
\includegraphics[width=15cm]{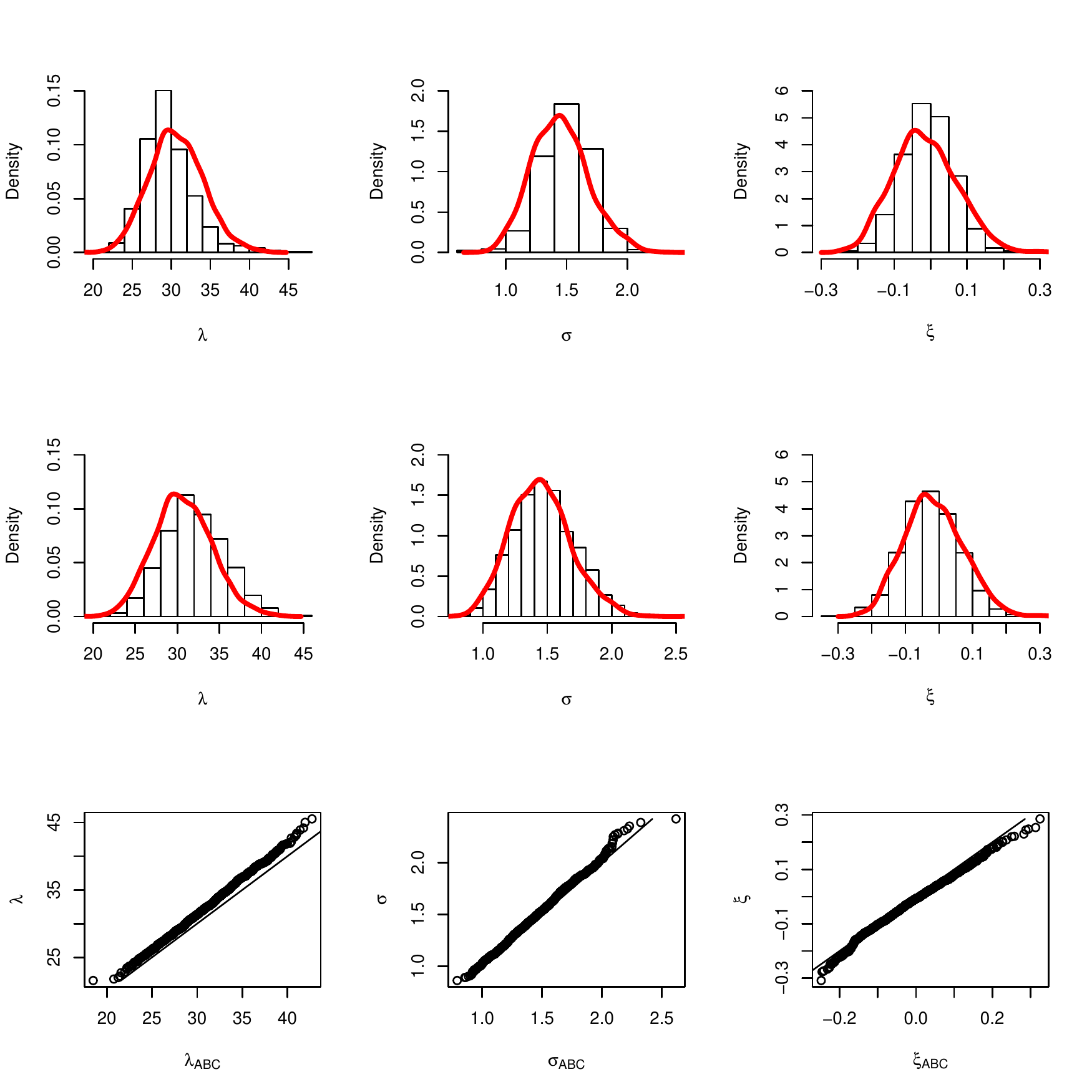}
\end{center}
\caption{  \label{MCMC-sphere}
Estimated marginal posteriors for the spherical inclusions model. Solid lines correspond to true marginal density, obtained  using standard MCMC. Histograms indicate marginal posteriors based on regression density estimation using: (top row) 112 summary statistics and (middle row) the 3 semi-automatic summary statistics (Fernhead and Prangle, 2012). 
Bottom row shows quantile-quantile plots of the 3 summary statistic approximation to the  true marginal distributions.
}
\end{figure}

The histograms in Figure \ref{MCMC-sphere} shows the regression density estimates of the marginal posteriors for the spherical inclusion model, based on 500,000 MCMC samples, obtained under the prior specification 
$\log\lambda\sim N(0, 100^2)$, $\sigma \sim Ga(0.01, 0.0001)$ and $\xi\sim N(0, 100^2)$. The top panels are based on using 112 summary statistics, with the middle panels using the 3 semi-automatic summary statistics.
The solid line indicates a density estimate of the true posterior marginal distribution derived from MCMC output using the known likelihood for the spherical inclusions model. The quantile-quantile plots compare the posterior estimated with the semi-automatic statistics  with the true marginals.
Even when constructing regression density estimates of the likelihood in 115 dimensions, reasonably good marginal parameter estimates are obtained, although they are clearly not perfect. The estimates improve considerably when only using 3 summary statistics, as the regression density estimation is then performed in only 6 dimensions. Clearly, the regression density approach has worked well, even in high dimensions.
Recall that these approximations are obtained based on flexible modelling with only 5,000 well placed samples $(\bS^{(i)},\btheta^{(i)})$ (though there are some computational overheads to identify where to place these). This compares favourably with earlier analyses of these data \shortcite{bortot+cs07} with the number of simulations in the tens of millions.
\\

\begin{figure}
\begin{center}
\includegraphics[width=15cm]{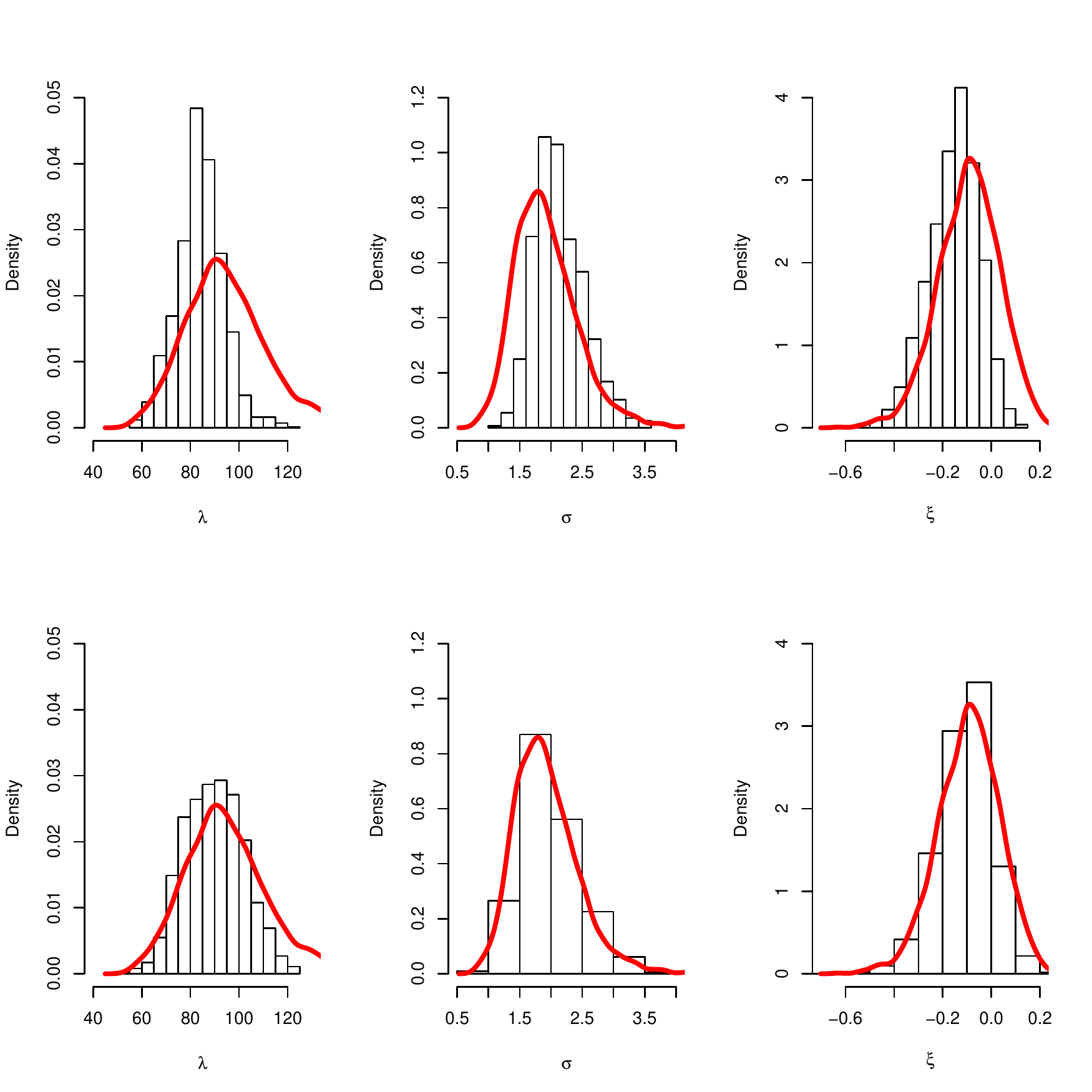}
\end{center}
\caption{  \label{elliphist}
Estimated marginal posteriors for the ellipsoidal inclusions model. Solid lines correspond to the marginal density obtained by the ABC-MCMC approach of Bortot et al. (2007), with kernel scale parameter $\delta=0.33$. Histograms indicate marginal posteriors based on regression density estimation using: (top row) 112 summary statistics and (bottom row) the 3 semi-automatic summary statistics.  
}
\end{figure}

The histograms in Figure \ref{elliphist} display the marginal posterior regression density estimates for the ellipsoidal model. Again, the top panels are based on using 112 summary statistics, with the bottom panels using the 3 semi-automatic summary statistics.
The solid lines represent the estimate of the posterior distribution obtained using the MCMC-based ABC method of \shortciteN{bortot+cs07} with kernel scale parameter $\delta=3.3$. Under the spherical model, this method is able to correctly reproduce the true posterior for $\delta=3.3$ (or lower). \shortciteN{bortot+cs07} then assumed that this approach could correctly reproduce the true posterior for the ellipsoidal model, although they were unable to verify their results by reducing $\delta$ further.
From Figure \ref{elliphist}, as with the spherical inclusions model, while the marginal posterior distributions under the regression density approach using the 112 summary statistics appear less accurate than when using the semi-automatic statistics, they still have broadly the right shape and location. Of course, because of the lower-dimensional regression density modelling, when using the 3-dimensional semi-automatic statistics, the resulting density estimates are closer to the MCMC-based estimate of \shortciteN{bortot+cs07}, albeit obtained with a far smaller computational overhead.
In summary, these results appear to broadly support the previous study of the ellipsoidal model by \shortciteN{bortot+cs07}, although there is some indication that the posterior marginals obtained by \shortciteN{bortot+cs07} may be too wide, especially for $\lambda$. This would be consistent with the regression density estimate's construction at $\delta=0$ compared with that of $\delta=3.3$ for \shortciteN{bortot+cs07}.
\\

As previously discussed, once the estimate of $L(\bS|\btheta)$ is obtained, it is then a simple matter to perform a range of likelihood-based inferences that are not immediately available with standard ABC methods.
Table \ref{table1} provides numerical estimates of the mean and the 0.025 and 0.975 quantiles of the 
marginal posterior for each parameter under various models and analyses, based on the semi-automatic summary statistics.
The estimates $(\lambda_{mle},\sigma_{mle},\xi_{mle})$ denote the maximum likelihood estimates based on $\hat{L}(\bS|\btheta)$, obtained by maximising the analytical expression for the approximate likelihood using the {\tt optim} function in {\em R}.
These frequentist estimates are consistent with  their Bayesian counterparts in Table \ref{table1} due to the uninformative prior implementation.
The estimates $(\lambda_{ABC},\sigma_{ABC},\xi_{ABC})$ indicate those obtained using the MCMC-based ABC method of \shortciteN{bortot+cs07} with kernel scale parameter $\delta=3.3$. These analyses use the 112-dimensional summary statistics. For the spherical model, this ABC posterior is known to coincide with the true posterior, and this is in evidence in the results, although the upper tail for $\lambda_{ABC}$ is estimated slightly shorter. As with Figure \ref{elliphist}, the marginal posteriors for the \shortciteN{bortot+cs07} analysis are wider than those obtained from $\hat{L}(\bS|\btheta)$.
\\

Finally, the estimates $(\lambda_{a},\sigma_{a},\xi_{a})$ correspond to those obtained under the regression density approach for different choices of the prior for $\xi\sim N(0,a^2)$, for $a=100, 1, 0.5$ and $0.1$. The original analysis of \shortciteN{bortot+cs07} with $a=100$ did not incorporate any information about the upper-tail heaviness of inclusion sizes: $\xi>2$ is physically unlikely, as this would result in extremely heavy tails; $\xi<-1$ is also unlikely as this would result in a very clear maximum inclusion size truncation. It is computationally trivial to implement this prior sensitivity analysis once the estimate $\hat{L}(\bS|\btheta)$ has been obtained, in contrast with standard ABC approaches. The sensitivity analysis in Table \ref{table1} indicates that the vague prior ($a=100$) does not affect the conclusions, with differences in the posterior only emerging for very strong prior beliefs ($a=0.1$).
\\

For both spherical and ellipsoidal inclusion models we additionally constructed a likelihood estimate by simply fitting a mixture of multivariate normals to the samples $(\bS^{(i)},\btheta^{(i)})$ and then conditioning on $\btheta$. This is equivalent to our regression density approach without the transformation $(\bS,\btheta)\rightarrow(\bU,\btheta)$, or the method of \shortciteN{bonassi11} but conditioning on $\btheta$ to obtain a likelihood, rather than on $\bS$ to produce a posterior. For both models this approach performed reasonably well when the 3 semi-automatic statistics were used (results not shown). However, it became numerically unstable in the 115 dimensional setting. That the regression density approach is numerically  stable even in high dimensions indicates the importance of the flexible mixture of experts transformation, $(\bS,\btheta)\rightarrow(\bU,\btheta)$, in this setting.

\section{Discussion}
\label{secDiscussion}

Approximate Bayesian computation using an estimated
likelihood function has many advantages over direct estimation of the posterior.
As such, we argue that methods that focus on likelihood construction should be considered
an important part of the ABC toolbox.
While conditional 
density estimation in high dimensions is challenging, it is an active area of current research.
As the reliability and flexibility of these techniques improve, this will only enhance the value of the likelihood-focused approach to ABC.

\bibliographystyle{chicago}
\bibliography{ABC-ME}

\begin{table}[hbt]
\begin{centering}
\begin{tabular}{|l|c|c|c||c|c|c|}
\hline
&\multicolumn{3}{c||}{{\bf Spherical}} &\multicolumn{3}{c|} {{\bf Ellipsoidal}}\\
\hline
  &0.025 &mean &0.975 &0.025 &mean &0.975\\
\hline
  $\lambda_{mle}$ &24.25&31.67&39.08&69.53&96.77&124.01\\
  $\lambda_{ABC}$&24.26&30.61&37.93&65.01&95.18&135.11\\ 
  $\lambda_{100}$ &25.00&31.59&39.29&68.82&93.40&119.38\\
  $\lambda_{1}$&25.09&31.65&39.34&69.41&93.66&118.28\\
  $\lambda_{0.5}$&25.05&31.63&39.32&70.42&93.97&116.76\\
  $\lambda_{0.1}$&25.18&31.77&39.44&72.04&94.98&117.84\\
  \hline
   $\sigma_{mle}$&1.01&1.48&1.94&1.08&1.84&2.60\\
  $\sigma_{ABC}$&1.02&1.45&1.95&1.12&1.92&3.06\\
  $\sigma_{100}$&1.03&1.46&1.95&1.08&1.88&2.96\\
  $\sigma_{1}$&1.03&1.46&1.96&1.07&1.87&2.95\\
  $\sigma_{0.5}$&1.03&1.46&1.95&1.09&1.85&2.86\\
  $\sigma_{0.1}$&1.07&1.44&1.86&1.11&1.68&2.34\\
  \hline
  $\xi_{mle}$&-0.21&-0.03&0.15&-0.29&-0.08&0.14\\
  $\xi_{ABC}$&-0.18&-0.02&0.17&-0.35&-0.09&0.14\\
  $\xi_{100}$&-0.18&-0.02&0.15&-0.33&-0.09&0.12\\
  $\xi_{1}$&-0.18&-0.02&0.15&-0.33&-0.08&0.12\\
  $\xi_{0.5}$&-0.17&-0.02&0.15&-0.31&-0.08&0.12\\
  $\xi_{0.1}$&-0.13&-0.01&0.12&-0.18&-0.04&0.10\\
  \hline
\end{tabular}
\caption{Posterior summaries for $\btheta=(\lambda,\sigma,\xi)$ for the spherical (left) and ellipsoidal (right) models. Columns correspond to the estimated 0.025 and 0.975 quantiles and the posterior mean for each parameter, under each model. The {\em mle} parameter suffix indicates results from maximum likelihood estimation (giving the MLE and lower and upper 95\% confidence limits), {\em ABC} indicates estimates from the ABC posterior using the approach of Bortot et al. (2007) with $\delta=3.3$. A numerical suffix (e.g. $\lambda_a, \sigma_a,\xi_a$) indicates the prior for $\xi$ is $N(0,a^2)$ under the regression density approach.
}\label{table1}
\end{centering}
\end{table}


\end{document}